\documentclass{aa}  
\newcommand{\filter}[1]{\mbox{\it #1\/}}

\usepackage{graphicx}
\usepackage{color}
\usepackage{hyperref}
\hypersetup{colorlinks=true,linkcolor=blue,citecolor=blue,filecolor=blue,urlcolor=blue}
\usepackage{tikz}
\usetikzlibrary{calc,positioning}
\usepackage{ulem}
\usepackage{subfigure}
\usepackage{ifthen}
\usepackage{placeins}
\usepackage{lipsum}
\usepackage{float}
\usepackage{xcolor}
\usepackage{xspace}

\usepackage{txfonts}

\definecolor{lime}{HTML}{A6CE39}
\DeclareRobustCommand{\orcidicon}{%
        \begin{tikzpicture}
        \draw[lime, fill=lime] (0,0) 
        circle [radius=0.16] 
        node[white] {{\fontfamily{qag}\selectfont \tiny ID}};
        \draw[white, fill=white] (-0.0625,0.095) 
        circle [radius=0.007];
        \end{tikzpicture}
        \hspace{-2mm}
}

\newcommand{\orcidMA}{\href{https://orcid.org/0000-0002-5306-4089}{\orcidicon}}
\newcommand{\orcidAZ}{\href{https://orcid.org/0000-0001-9509-7316}{\orcidicon}}
\newcommand{\orcidJK}{\href{https://orcid.org/0000-0002-8762-7863}{\orcidicon}}
\newcommand{\orcidSB}{\href{https://orcid.org/0000-0002-9091-2366}{\orcidicon}}
\newcommand{\orcidFC}{\href{https://orcid.org/0000-0003-2658-7893}{\orcidicon}}

\begin{document}

\title{Young star clusters in the extended UV disk of M83}
    \subtitle{Star formation in a low-density, low-metallicity environment}

   \author{M. Andersen
          \inst{1}\orcidMA
          \and
          A. Zavagno\inst{2,3}\orcidAZ
          \and 
          J. Koda \inst{4}\orcidJK
          \and S. Boissier 
          \inst{2}\orcidSB
          \and 
          F. Combes. 
          \inst{5}\orcidFC
          }

   \institute{European Southern Observatory, Karl-Schwarzschild-Str. 2, 85748 Garching, Germany\\
     \email{morten.andersen@eso.org}
         \and
        Aix Marseille Univ, CNRS, CNES, LAM, Marseille, France 
   \and 
   Institut Universitaire de France, 1 rue Descartes, 75005 Paris, France 
     \and 
     LERMA, Observatoire de Paris, Coll`ege de France, PSL University, CNRS, Sorbonne University, Paris
            \and 
                Department of Physics and Astronomy, Stony Brook University, Stony Brook, NY 11794-3800, USA;\\}

  \abstract
   {The process of star formation in a low-metallicity environment and whether it differs from star formation in the solar neighbourhood is an open topic. { Recently, CO clumps were identified within the extended ultraviolet (XUV) disk of M83 despite the low metallicity and distance from the galaxy center which allows a search for any recent star formation in the clumps.}}
  { We aim to probe whether active star formation is occurring within the CO clumps in the XUV disk, and to assess its implications for star formation in such environments.}
   {We analyse deep Subaru Suprime-Cam H$\alpha$, Hyper Suprime-Cam g, r, \& i, and Gemini/FLAMINGOS 2 $J$ and $Ks$ band imaging to search for evidence of star formation within the ALMA-identified CO clumps. We identify sources within the  CO clump contours, and we evaluate the probability that these sources are associated with the CO clumps rather than being chance alignments. Based on their magnitudes and adopting single-age stellar population models, we estimate their masses. }
   {Four of the clumps have an associated near-infrared counterpart. Based on their magnitude, color, and the general source density in the field, two of these could be chance alignments. One near-infrared counterpart is associated with the most massive clump in the sample, while another is an unresolved cluster near-infrared source  with an estimated age of 5~Myr and a mass of 2000~M$_\odot$. The other, an extended complex, is consistent with an age of 6~Myr and a mass of 700~M$_\odot$. The near-infrared photometry is sensitive to young clusters with masses down to a few hundred of M$_\odot$ for a fully sampled IMF. }
   {Star formation is ongoing in a few of the CO clumps, and despite their location on the outskirts of M83, favourable conditions for star formation persist in at least a few clusters. One of the detected sources appears to be interacting with an adjacent molecular clump, which could trigger further star formation.}

   \keywords{ Galaxies: star formation -- Galaxies: star clusters: general}

   \maketitle
\section{Introduction}
The impact of early (pre-supernova) radiative feedback from massive stars on future star formation (SF) can be either \textit{positive} (by triggering SF through the gravitational collapse of molecular clump) or \textit{negative} (by quenching SF through the removal of the gas). In our galaxy, the early radiative feedback of massive stars, along with the compression of their associated ionised (H\,{\sc{ii}}) regions, shapes the surrounding interstellar medium and can trigger the formation of a new generation of stars, including massive ones \citep{2010A&A...523A...6D}. This phenomenon has also been observed in the Large Magellanic Cloud \citep[e.g.]{2016A&A...592A..77B}. Recent results obtained with the James Webb Space Telescope on a sample of nearby galaxies (the PHANGS sample, Physics at High Angular resolution in Nearby GalaxieS, \citet{2019Msngr.177...36S}) show the central role of early radiative feedback in shaping the surrounding interstellar medium and deeply influencing the star formation laws \citep{2023ApJ...945L..19S,2023ASPC..534....1C,2024ApJS..273...13W}. { The role of H\,{\sc{ii}} regions can be further explored in different environments, such as the low-density, low-metallicity extended ultraviolet (XUV) disks of galaxies, where star formation is not expected to occur based on the previously low success of detecting molecular clouds \citep{watsonkoda}, and the long time scale for converting the gas into stars in the outer disk of M83. 
Near-infrared imaging can be particularly useful to detect young stellar clusters due to the intrinsic brightness at longer wavelengths of pre-main sequence stars and the reduced effects of extinction. 
This will help us to understand the possible environment dependence  on star formation}.  \\
NGC~5236 (Messier 83, hereafter M83) is a one of the nearest barred spiral galaxy, located at a distance of only 4.5 Mpc \citep{2003ApJ...590..256T}. It has been extensively studied at many wavelengths \citep[e.g.][]{sextl2025,russell2020,foyle2012,boissier2005}, but it is only with GALEX that the presence of a large disk beyond $\sim$ 5 kpc was found \citep{2005ApJ...619L..79T}. This extended disk was then further studied to find and study H\,{\sc{ii}} regions \citep[e.g.][]{gildepaz07,bresolin09}. 
Atomic gas was mapped to large radius \citep[e.g.][]{bigiel2010,heald2016,eibensteiner2023}. Molecular gas in the extended disk eluded ALMA observations of the CO $J=2-1$ transition \citep{bicalho}, but was finally detected  in CO(3-2) \citep{koda22,koda2024b}.

M83 geometry being almost face-on, it allows a non-obstructed view of the disk of the galaxy. Subaru observations reveal the presence of faint H{\,\sc{ii}} regions in its outskirts, outlining recent SF involving early-type stars \citep{koda2012}. ALMA CO(3-2) observations \citep{koda22,koda2024b} towards part of the faint XUV disk of M83 reveal a series of 23 CO compact regions (each smaller than the resolution of 6-9~pc), clustering within a small 1~kpc$^2$ region, whose spatial distribution seems to be tightly linked to that of the faint ionized regions observed in the field with Subaru (see Figure~\ref{overview}). For example, H{\,\sc{ii}} region "G" in \citet{koda22} and the surrounding ALMA CO(3-2) clouds \#12 to \#15, \#17 and \#20 are suggestive of a potential connection between the H{\,\sc{ii}} region and the clouds traced in CO (see Figure \ref{cloud_mos}). However, the origin and fate of these compact high-density clouds are unknown. Their apparent tight spatial relationship with H{\,\sc{ii}} regions suggests that these clouds might have been formed by the interaction of the expanding ionised gas with the surrounding medium, as observed in the galaxy \citep{2020A&A...638A...7Z} and the LMC \citep{2019ApJ...886...14F}, indicative of a \textit{positive} feedback. They could also be the result of gas expulsion after the recent star-forming process has consumed the gas and pushed the remaining gas away, suggestive of a \textit{negative} feedback.  \\
GALEX observations revealed that the Ultra Violet (UV) emission extends well beyond the optical disk of galaxies \citep{gildepaz2005,thilker2007}. The eXtended UV (XUV) disks emission suggests that star formation can occur in a larger fraction of the H\,{\sc{i}} disk, up to 4 times the optical radius \citep{bicalho}. However, in the XUV disks the Star Formation Rate (SFR) is observed to be lower than in the central galaxy disk \citep[e.g.][]{bigiel2010}, suggesting that positive feedback might dominate at the center while negative feedback might dominate outside. 
Alternatively, the star formation efficiency could be lower due to low pressure \citep[e.g.][]{shi2011}. A way to distinguish between these two scenarios is to look
at the properties of CO clouds, with respect to the H\,{\sc{ii}} regions
and young star clusters associated with the clouds. The latter involved deep panchromatic optical and near-infrared observations
to detect clusters to low masses and through extinction if the
clusters are still embedded in the molecular clouds. To this end,
we present deep Gemini/FLAMINGOS-2 (F2) J and Ks band
imaging together with Subaru Hyper Suprime-Cam gri imaging
of the CO clumps detected in the XUV disk of
M83. 
Combination of far UV observations, (tracing the old star formation, $\sim$ 100 Myr), H$\alpha$ (tracing the recent $\sim$ 10 Myr star formation) and dense molecular tracers such as CO $J=3-2$ (hereafter CO(3-2)) and their possible near-infrared counterparts (tracing the young star formation) allows us to map the star formation history and to put new observational constraints on feedback processes.  \\ 

The paper is organised as follows: Section~\ref{data} presents the data used and the data reduction. The results are presented in Section~\ref{res} and discussed in Section~\ref{disc}. The summary of the main findings and conclusions is given in Section~\ref{conc}.   

\section{Data and data reduction} \label{data} 
GALEX showed UV emission well beyond the optical disk of M83 which suggested that star formation was still ongoing. 
\citet{koda22} presented ALMA $CO(3-2)$ observations of a $\sim$0\farcm75$\times$ 0\farcm85 field of view of the UV disk of M83 at a distance of 7\farcm8 from the center of M83, corresponding to 10~kpc. 

We have recently obtained deep near-infrared imaging in the \filter{J} and \filter{Ks} bands, using FLAMINGOS 2 mounted on Gemini South. 
The observed field is centered on the $\sim$0\farcm75$\times$ 0\farcm85 ALMA pointing in \citet{koda22}, in order to identify and characterize any star clusters associated with the CO clumps identified in \citet{koda22}. 
These observations are combined with optical SUBARU Hyper Suprime-Cam (HSC) observations in the gri bands together with SUBARU Suprime-Cam H$\alpha$(NA656\footnote{https://www.naoj.org/Archive/Instruments/SCam/sensitivity.html}) and R band observations. 

\begin{figure}[h] 
\includegraphics[width=9cm]{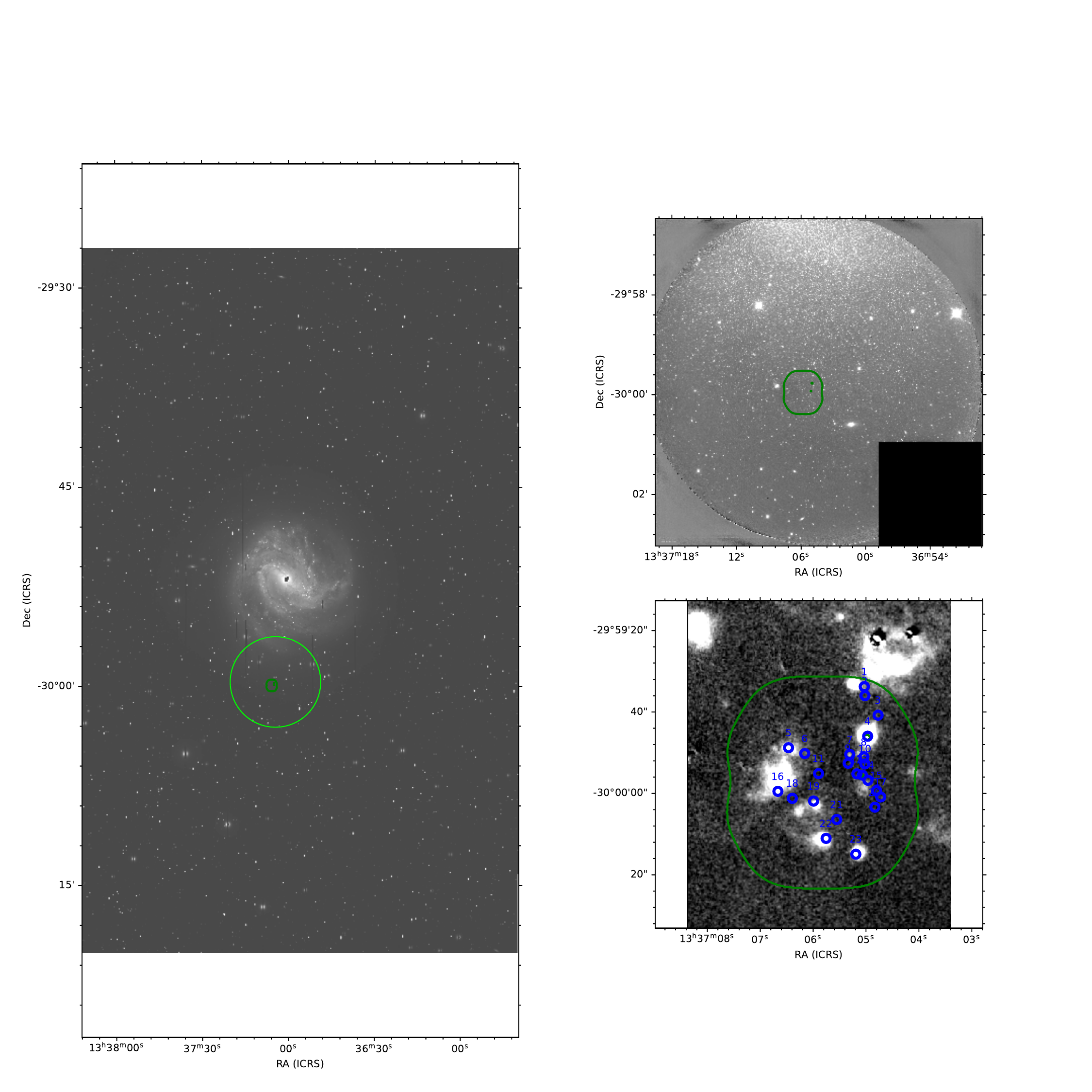}
\caption{Left: Subaru Suprime-Cam $R_C$ band image of M83 \citep{koda2012}. Shown are the field observed by ALMA (green oval contour), and the F2 field of view (green circle). Upper right: F2 Ks band image. The ALMA field is marked with the green oval contour.  Lower right: Continuum subtracted Subaru Suprime-Cam H$\alpha$ image of the 1.1$\times$1.5\arcmin field of view around the ALMA field, shown as the green oval contour. The CO clumps identified with ALMA are indicated in blue. }
\label{overview} 
\end{figure}

  Fig.~\ref{overview} shows the Subaru R image of M83 with the region observed with ALMA marked with a green oval shape, and the location of the ALMA clumps in blue, together with the indication of the radius of the optical disk.

\subsection{Gemini F2 data}
We have observed the region around the ALMA field using F2 on Gemini South, program GS-2022A-FT-204 (PI Koda). 
The field of view is  6\arcmin\ diameter with a pixel scale of 0.18\arcsec . 
The observations through the J and Ks bands were obtained on April 18, 2022 under clear conditions. 
Guiding was performed using the peripheral wavefront sensor, causing vignetting in the south-western corner that is masked in the final images. 
Observations were performed with an on-off pattern with 6 integrations on source followed by three dithered sky integrations. 
Individual exposure times were 30 seconds in the J band and 12 seconds in the Ks band, respectively. 
A total of 63 exposures in the J for a total exposure time of 1890 seconds and a total of 203 on-source integrations in the Ks band for a total exposure time of 2436 seconds were obtained. 

\begin{figure*}
   \centering
   \includegraphics[width=9cm]{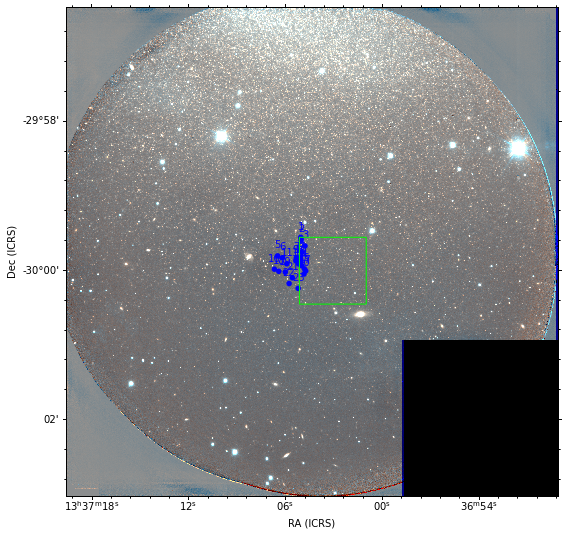} 
   \includegraphics[width=9cm]{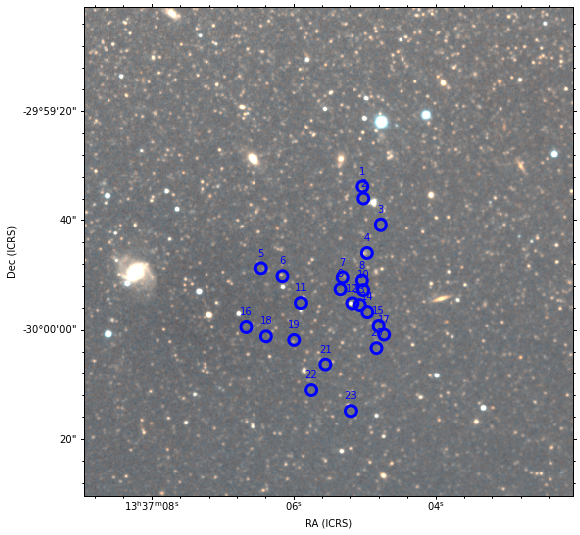}
  
      \caption{Left: Color-image of the Gemini F2 observed region. J band is the blue channel, Ks is the red channel and average of the J and K bands is used for the green channel. The blacked out area to the south-west is where the guide arm is located. The green box marks the control region used to calculate the source density.
      Right: Zoom of the region observed by ALMA. The color scale is the same as in the left figure. The CO emitting regions identified by \citet{koda22} are shown as blue circles.  The center of M83 is to the north which is the reason for the higher stellar density in that direction.
      }
         \label{col_image}
   \end{figure*}

The exposures were reduced in a standard manner using the DRAGONS pipeline \citep{2019ASPC..523..321L}. 
Briefly, the individual frames were dark subtracted before flat-fielding, using the darks and flats closest in time available in the archive. 
A running sky frame using the most recent sky frame positions was created to subtract from the science exposures.
Individual science exposures were registered and co-added for each filter. 
The final spatial resolution (FWHM) is 0$\farcs$47 in the $J$ band and 0$\farcs$38 in the $K_{\rm{s}}$ band, respectively. 
This corresponds to a physical scale of $\sim$10~pc at the distance of M83. 
The final \filter{J}  and \filter{Ks} band images were aligned to GAIA DR3 \citep{gaiadr3} using the astronomy package by \citet{lukas}. 
Fig.~\ref{col_image} shows the Gemini F2 observed field with the location of the observed CO clumps overlaid.

\subsection{ALMA and SUBARU observations} 
The ALMA data are discussed in more detail in \cite{koda22}. Briefly, the CO(3-2) line was observed over a 0$\farcm$750 $\times$ 0$\farcm$850 region, corresponding to 0.98$\times$1.11~kpc$^2$. 
Some 23 clumps were identified by \cite{koda22} with estimated gas masses ranging from 820 M$_\odot$ to 8700 M$_\odot$. 

\citet{koda2012} presented deep R, and H$\alpha$ imaging of a large part of M83 on September 7, 2010 using Suprime-Cam on the Subaru telescope. 
Data were reduced in a standard manner and have a 1 $\sigma$ sensitivity of 24.90 and 25.56 AB mag in the H$_\alpha$ and $R_C$ bands, respectively. 
These observations were later complemented with HSC imaging in the g,r, and i filters, similar to the SDSS filters of the same name, on March 20, 26, and 27, 2015.
The data were calibrated in a standard manner using the hscPipe package (version 8.5.3) \citep{bosch2018}.

\subsection{Source detection and photometry}
Source detection and photometry were performed using the {\tt daophot} package \citep{stetson} implemented in {\tt iraf}. 
Sources were identified in the J band down to a detection 4$\sigma$ above the background level. This source list was then used for photometry in both the J and $Ks$ bands. 

Photometry was performed using a source aperture of three pixel radius (0\farcs 54) to match the FWHM of the point sources in the images,  and using a sky annulus between 15 and 20 pixels. 
Sources were considered real if their center positions were the same in the two bands to within two pixels in RA and DEC, with a combined photometric error of less than 0.2 mag. 
{ To summarize the requirements for the near infrared catalogue: Objects down to a detection limit of 4$\sigma$ in the \filter{J} band with forced photometry in the \filter{Ks} band were considered for the catalogue if their positional match agreed within 2 pixels in both directions and a combined photometric error less than 0.2 mag.}
 A total of 11055 sources fulfill these criteria. 

\begin{figure*}
\includegraphics[width=18cm]{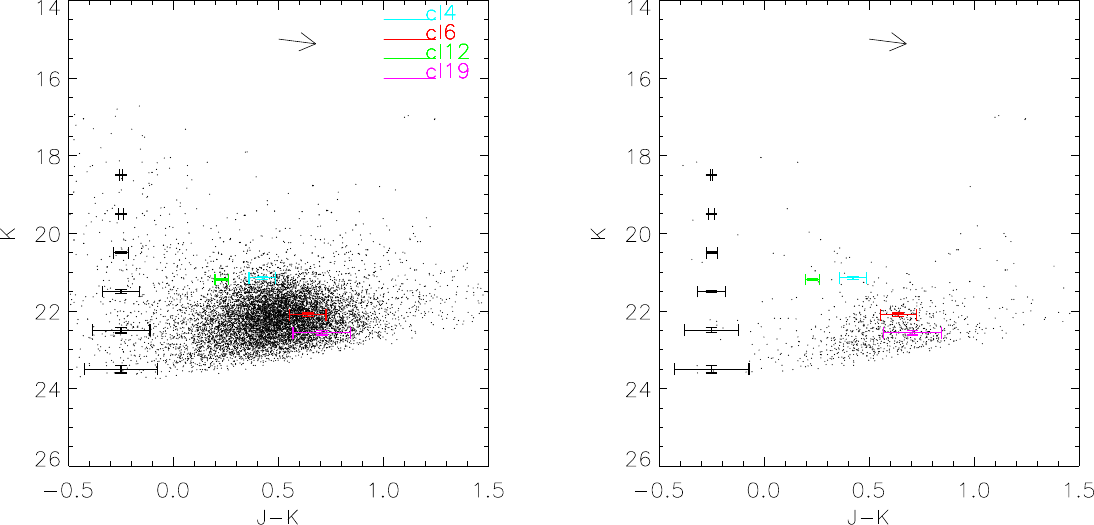}
\caption{Left: The J-K$s$ vs K$s$ color-magnitude diagram of the F2 field around the ALMA pointing and the reference field. Typical error bars as a function of K band magnitude are shown to the left in the plot. 
 The four sources associated with ALMA clumps are marked and their error bars are included. 
A reddening vector corresponding to the effect of $\mathrm{A_V}=1mag$ is shown. 
Right: Same as above but only for the sources within the ALMA field of view and the control region shown in Fig.~\ref{col_image}.}
\label{CMD}
\end{figure*}

Figure~\ref{CMD} shows the resulting color-magnitude diagram for all sources identified in both J and Ks bands. 
The sources spatially coinciding with ALMA CO clouds as quantified below are marked in red.  

The final source list from the near-infrared photometry was then used as input for photometry in the optical g, r, and i bands. 
An aperture of 6 pixels (0\farcs 9) radius was used to match the seeing in the visible data and the instrumental photometry was zeropoint calibrated to the PAN-STARRS \citep{chambers2016} photometry using stars in common between the two datasets.

The FLAMINGOS-2 photometry was placed in the 2MASS systems using 2MASS point sources \citet{cutri2003} that were not saturated in the F2 images and calculating the zero point for both the J and Ks band observations. The 2MASS magnitudes were then converted to AB magnitudes using the conversions in \citet{blanton}.

\section{Results} \label{res}

\subsection{Sources aligned with the clumps}
\begin{figure} 
\includegraphics[width=9cm]{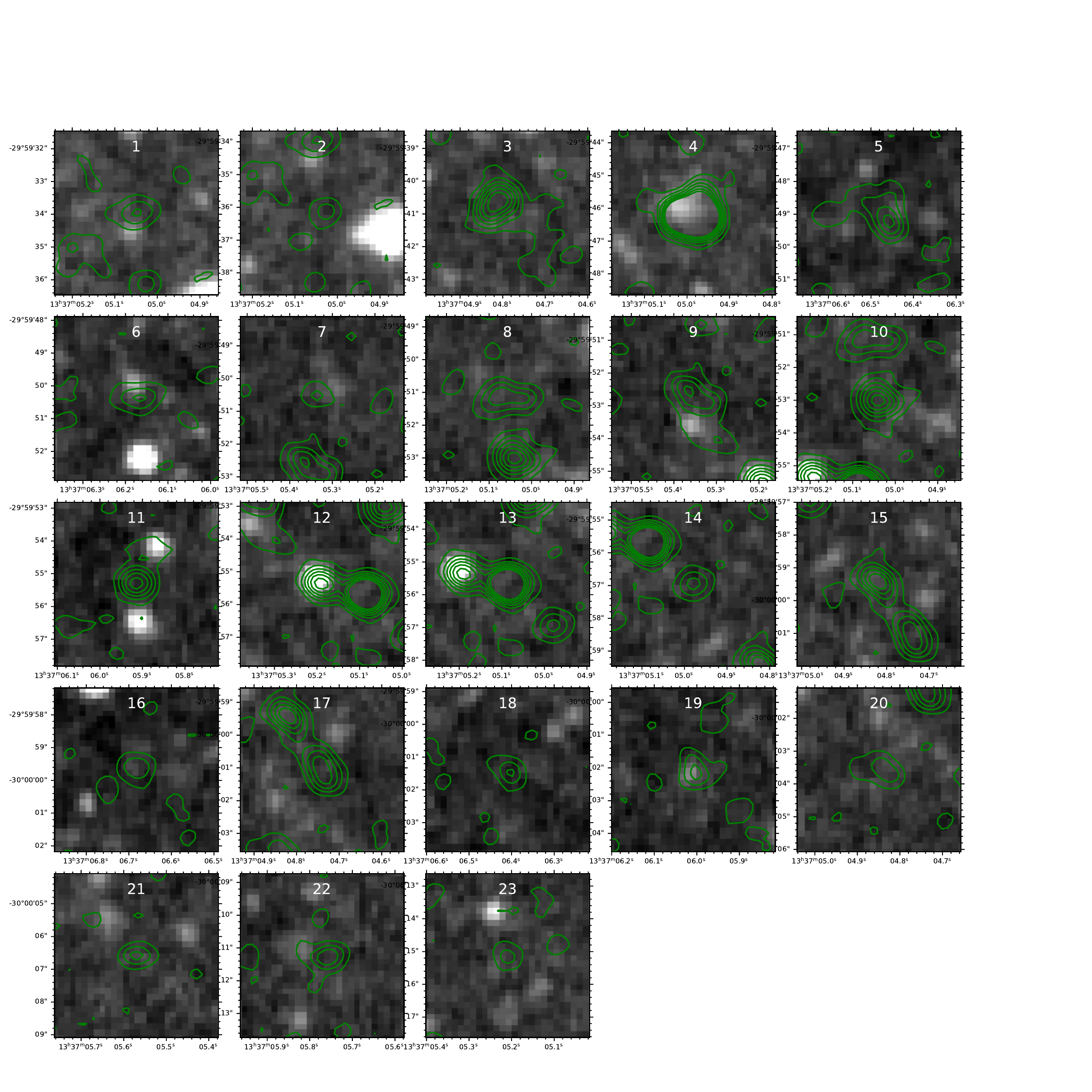}
\caption{The central 5\arcsec\ in the Ks band around each of the CO clumps identified in \citet{koda22}. For the ones with a  identification of a near-infrared source within 1.5$\times$ FWHM   from the center of the CO clump. the photometric parameters are provided in Table~\ref{counterparts}. Contours are the zero moment CO emission, Each postage stamp is labelled with the number of the CO clump identified in \citet{koda22}. }
\label{zoom_CO}
\end{figure}

Fig.~\ref{zoom_CO} shows a 5\arcsec\ Ks band postage stamp around each of the CO clumps identified in \citet{koda22}. 
The green contours show the CO emission within each box. 

\begin{figure}[h] 
\includegraphics[width=9cm]{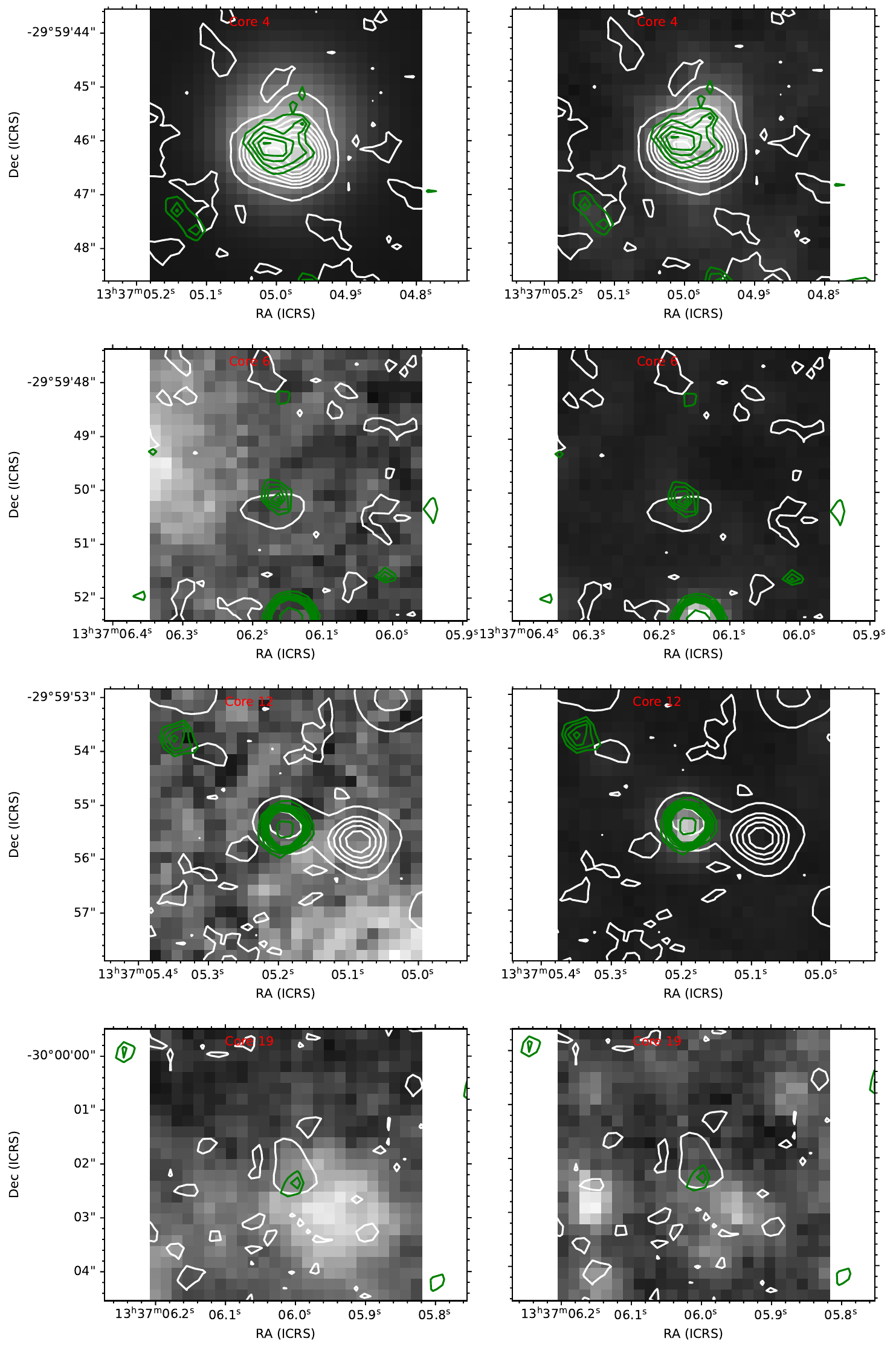}
\caption{Left column: H$\alpha$ postage stamp of a 5\arcsec\ square around the center of the molecular clouds associated with a near-infrared source, i.e. cloud 4, 6, 12, and 19. 
The J band flux is overplotted as green contours and white contours are the CO(3-2) intensity contours. Right column, the Subaru i band image shown with the same contours as in the left column.}
\label{cloud_mos} 
\end{figure}

Our criterion for a source to coincide spatially with a CO clump, is to be within a radius of 1.5 times the FWHM of the CO clump (averaged over the x and y widths) from the center of the CO clump as determined in \citet{koda22}.
Of the 23 CO clumps identified in \citet{koda22} we find a good spatial alignment with a near-infrared source in four cases; clump 4, 6, and 12, and 19. 
Their photometry is presented in Table~\ref{counterparts}. 
A comparison of the emission at several wavelengths for the four clumps is shown in Fig.~\ref{cloud_mos}.
Choosing a different distance criteria will slightly change the results. Using a tighter requirement of 1.25 times the FWHM as the threshold results in three alignments, clumps 4, 12, and 19. 
Conversely, a threshold of 1.75 times the FWHM the number of alignments is still four sources. 

{ The main limitations in the sample are the sensitivity and potential reddening of the star clusters associated with the molecular clouds. The combined error of 0.2~mag means that likely some objects are not identified due to their low signal to noise and hence the probability of a detected object associated with a molecular cloud could be estimated to be lower than a fully complete sample would suggest. This is mainly the case for the source associated with cloud 19, where a stricter combined magnitude cut of e.g. 0.1 mag means the object would not have been in the catalogue. 
The remaining sources are substantially brighter and less affected by any incompleteness. }

We briefly discuss the morphology of each source associated with a CO cloud in the following. We further evaluate the reliability of the physical association of the source with the clump. 

With over 11000 near-infrared sources detected within the Flamingos-2 field-of-view, there is a substantial chance that a given clump may be associated with a near-infrared by coincidence. 
The full near-infrared color-magnitude diagram allows an estimate of the probability that a star associated with an ALMA clump is a chance alignment based on the brightness of the object. 
{ We estimate the density of field objects from a 300$\times$300 pixel box adjacent to the ALMA field, shown as the green square in Fig.~\ref{overview}. The probability of an object being a field object is then estimated through the density of objects in the control field at least as bright in the Ks band as the cluster candidate and like for the cluster candidates detected in both the \filter{H} and \filter{Ks} bands. This density is then multiplied by the search radius around the center of the CO cloud based on the FWHM of the core. 
The probability of the object being a field object is then calculated using Poisson statistics (the probability of having at least one object brighter than the candidate cluster). 
 
We have repeated the membership estimate calculation using three different control fields of the same size but shifted in RA, ensuring they are at a similar radial distance from the center of M83 as the ALMA field. 
 The probability of detecting a star of the observed magnitude or brighter for each source associated with a molecular clump is provided in Table~\ref{counterparts}, where the quoted uncertainty is the range estimated from the three control fields. }
 
\subsection{{Properties of individual clumps}}
{ Although individually the probability of the source being aligned by chance is relatively small, given that there are 23 clumps, the cumulative probability for a chance alignment is not negligible. 
We have estimated this using the binomial distribution and for using the probability of 5\%\ for a chance alignment. 
The probability of at most one source by chance is then 69\%\ and the probability for at most two sources is 90\%.

We have further tested the frequency of chance alignments by randomly positioning the CO clumps within the ALMA field of view and determining the probability of chance alignments. We find the chance of finding at most 1 chance alignment to be 58\%\ and 82\%\ for at most two chance alignments, comparable to the values from the binomial distribution. }

We discuss below the clumps with a near-infrared source nearby. 
{\bf Clump 4}: 
The source associated with clump 4 is relatively faint and red. 
Looking in detail, it is slightly spatially extended in both the \filter{J} and \filter{Ks} band images. 
The probability of the source being a chance alignment relatively low at one percent and the fact that it is extended further suggests it is physically associated with the molecular clump.

The optical images further show an extended structure, but only partially overlapping with the peak of the near-infrared emission. Further comparison with Subaru H$_\alpha$ imaging shows that the CO clump coincides with the brightest H$\alpha$ region within the ALMA field of view.  
A curve of growth of the near-infrared source centered on the peak of the optical emission suggests that the emission extends over a one arcsecond radius, similar to what is seen in the optical but at lower spatial resolution. The wide-scale near-infrared emission thus also covers the visible emission but without any clear structure, at least partly due to the high background in the near-infrared. 
The extended nature suggests a size of more than 20~pc, which in turn suggests that this may be part of a star-forming complex at the distance of M83. 
{ As a likely stellar complex instead of a single cluster it is very possible that the system is not described by a single age, which we discuss further in Sec. 4.}

The H\,{\sc{ii}} region associated with the clump has previously been identified in \citet{gildepaz07} and the electron temperature and metallicity estimated in \citet{bresolin09}. 
They determined an electron temperature of $T_e=11810 \pm640$K and $c(H\beta)$ of 0, which implies little to no extinction between us and the cluster. 
This further suggests that the H$\alpha$ emitting region is in the foreground of the molecular cloud. 

Adopting the center of the cluster from the SUBARU photometry, we have forced the center of the near-infrared emission to be the same and integrated the flux in all wavelengths for a 1.5\arcsec\ aperture to obtain the total flux of the extended cluster. 
The photometry is included in Table.~\ref{counterparts}. 

{\bf Clump 6:} The near-infrared source is located just to the north of clump 6 as indicated in Fig.~\ref{zoom_CO}. 
The point source is also identified in the Subaru i band, but not at shorter wavelengths.

{\bf Clump 12:} This relatively bright point source is located on the peak of the CO clump. It is also visible in the Subaru broadband data. However, there is no H$\alpha$ emission associated with the region. 
Due to the brightness, the probability of a chance alignment is low, as noted in the Table~\ref{counterparts}. 
The clump is part of what appears to be a string of CO clumps covering $\sim$ 300 pc.
Yet, this is the only clump with an infrared source associated with it. The string of clumps appear to be pointing towards the center of M83.

{\bf Clump 19:} The source associated with clump 19 is faint, only barely detected in the \filter{J} and \filter{Ks} bands. There is no source associated with it in the Subaru images. The peak of the diffuse H$\alpha$ emission is offset from the near-infrared source. The probability of the object being spatially aligned by chance is relatively high. However, if it is not a chance alignment, the very red color suggests a deeply embedded object. 
The offset position of the diffuse H$\alpha$ emission relative to the near-infrared source may be due to the source's  embedded nature, which would render its optical emission undetectable. 
However, given the relatively low mass of the molecular clump as estimated from the CO emission would overall suggest a chance alignment. 

In addition to the clumps with near-infrared sources associated with them, there is some evidence for a cluster associated with clump 7. 
clump 7 is associated with a very weak emission and a point source in the Subaru i band with a tentative detection in the J and Ks band but below the detection limit of 4$\sigma$ in either band. 
Thus, this might be a low-mass cluster at the limit of detection, as discussed below, of a few 100 M$_\odot$. 

Thus, in total 5 of the clumps are associated with a near-infrared source or a tentative source; two of those are possible chance alignments. 

\begin{table*}
\caption{Subaru and Gemini photometry of the near-infrared sources associated with CO clumps as identified in Fig.~\ref{zoom_CO}.  clump four has an optical source associate with it but it is not overlapping with the near-infrared source discussed here. clump 4 (ext) is the region associated with the HII region as discussed in the text.  }             
\label{counterparts}      
\centering                          
\begin{tabular}{c c c c c c c  }        
\hline\hline                 
CO clump ID & $G$ & $R$ & $i$ & $J$ & $Ks$ & Prob.\\    
\hline                        
       4 & - & - & - & 22.19$\pm$ 0.04 & 21.69$\pm$ 0.03&0.03$\pm$0.01\\
       6 & - & - & 24.48$\pm$ 0.25 & 22.74$\pm$ 0.07 & 22.10$\pm$ 0.04&0.04$\pm$0.01\\
      12 & 23.64$\pm$ 0.17 & 22.55$\pm$ 0.09 & 22.10$\pm$ 0.07 & 21.43$\pm$ 0.02 & 21.20$\pm$ 0.02&0.01$\pm$0.01\\
      19 & -  & - & - & 23.27$\pm$ 0.12 & 22.56$\pm$ 0.06&0.07$\pm$0.01\\
       4(ext) & 20.50$\pm$ 0.04 & 20.75$\pm$ 0.04 & 21.28$\pm$ 0.05 & 21.57$\pm$ 0.05 & 21.14$\pm$ 0.03 & -\\
      \hline                                   
\end{tabular}
\end{table*}

\section{Discussion} \label{disc}
\subsection{Cluster ages and masses}
Of the 23 CO clumps identified by \citet{koda22} we find a good spatial alignment of a near-infrared source for four of them. 

For the two sources with optical and near-infrared photometry, we can estimate the cluster masses assuming a single stellar population. 
To this end, we have utilized the Starburst99 V7.0.1 models \citep{leitherer} adopting a metallicity of Z=0.008 and as a fiducial case a \citet{kroupa2001} Initial Mass Function from 0.1 to 100 M$_\odot$. 
The stellar track of Geneva assuming high winds were used. 
The effects of the winds are strongest at wavelengths below ~350nm and for solar metallicity models, and thus not in the wavelength range studied here. \citet{Leitherer14} shows an increase in the longer wavelength flux of $\sim 10\%$ over the non-rotating models.

The H$\beta$ line was computed using the Lyman photon flux QLyc from Starburst99 and a case B recombination with an electron temperature of 10$^4$ K \citep{osterbrock2006}.
Other lines were added assuming the same line ratios that are implemented in the Le Phare code \citep[Ilbert, private communication, see also][]{saito2020}, following \citet{schaerer2009}. Then, the magnitudes were computed by integrating the spectra whith the transmissions taken from the FLAMINGOS2 webpage\footnote{https://www.gemini.edu/instrumentation/flamingos-2/components}.
A subsolar metallicity is appropriate for the region observed based on the metallicity gradient measured by \citet{hernandez}, where the metallicity extrapolated from their equations 8-10  to the distance of the ALMA field from the galaxy center (1.2$R_{25}$) is found to be between $[Z]=-0.52$ and $  -0.24$. 
This is somewhat higher than what was found for the H\,{\sc{ii}} region associated with clump 4 in \citet{gildepaz07,bresolin09} of $12+log(O/H)=8.05\pm0.07$ but generally consistent with their findings at the radial distance of the ALMA field. 
We note that for young clusters, at the age expected for clusters associated with CO clouds, the effects of a different metallicity in our photometric bands are relatively modest for ages below 6 Myr, which the low-metallicity tracks being bluer in color by a few percent compared to solar metallicity. 

\begin{figure}[h] 
\includegraphics[width=8cm]{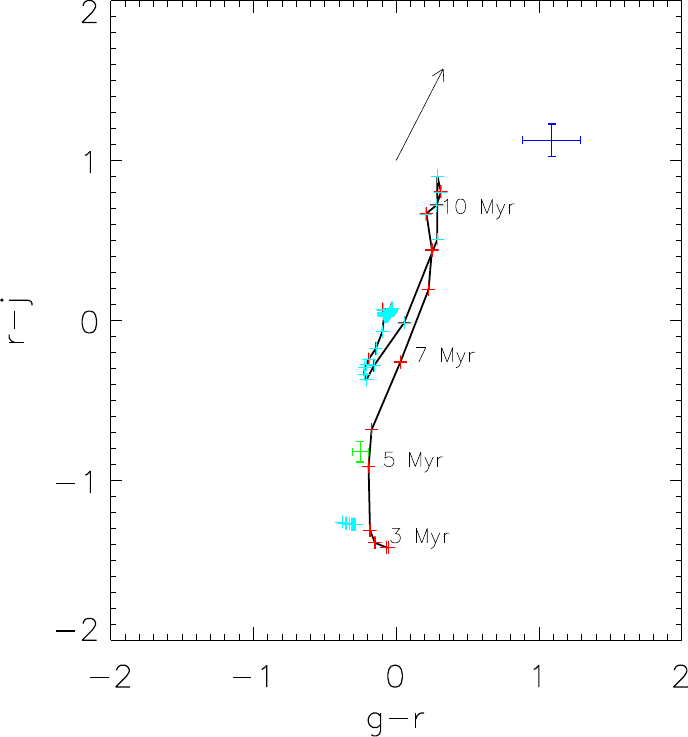}
\caption{ The g-r versus r-j color-color diagram for the source associated with cloud 4 (green) and 12 (blue). The aperture used for the extended cloud 4 source is 1.5 arcsecond. 
The Starburst99 isochrones for Z=0.008 and an upper mass limit of 100 M$_\odot$ (black line and red crosses for an age less then 50~Myr and for an upper mass limit of 20 M$_\odot$ (cyan points). 
The reddening vector corresponding to A$_\mathrm{V}=1$ is shown adopting the extinction law from \citet{cardelli} which was based on observations of early-type stars. We assume $R_V=2.75$, the average of  the LMC supershell and the SMC. 
Ages for the upper mass of 100~M$_\odot$ evolutionary track are marked. For the 20~M$_\odot$ track the lower left points are all in the 1-9 Myr range until the first giants appear. 
}
\label{CCDs} 
\end{figure}

Fig.~\ref{CCDs} shows the color-color diagram with the sources associated with clumps 4 and 12, which are the two sources with photometric measurements in both the visible and the near-infrared, with the Starburst99 isochrone overplotted.  
Both sources can be traced to the isochrone following the reddening vector for a single stellar population with an upper mass of 100 M$_\odot$. 
For clump 4 the reddening is within the error bars consistent with 0 (see the overlap with the isochrone in Fig.~\ref{CCDs} and the determined age is 6~Myr whereas for clump 12 the reddening is  A$_\mathrm{V}$=3.9 and the age is 4.5~Myr. 
The corresponding masses are 700 and 2400~M$_\odot$, respectively. 
These results are consistent with their spatial offset and overlap with the gas clumps and the presence and absence of associated H$\alpha$ emission, as well as moderate extinction for clump 4 and no extinction for clump 12. 

The IMF is a stochastic entity and the Starburst99 calculations assume a fully sampled IMF. For lower-mass clusters this is typically not a realistic assumption, and there will be substantial scatter owing to the poor sampling of the upper-mass part of the IMF. To illustrate this, we show in Fig.~\ref{CCDs} the evolution of a stellar population with an upper mass limit of 20~M$_\odot$ (cyan points). 
Due to the lack of very massive stars, the colors remain mostly constant for the first $10^7$ yr after which the evolution is very similar to the higher upper mass evolutionary track. 
{ Further, because of the steep luminosity function of massive stars with mass, the integrated magnitude per unit mass in the lower-mass cutoff models is 2.5 magnitudes fainter than in the case of a 100 M$_\odot$ upper-mass. 
This is the main uncertainty in the mass estimate.}
However, \citet{koda2012} argues that there is no indication of a truncated IMF for the XUV disk in M83 and thus the limits of a truncated IMF are likely an upper limit to the effects of the stochasticity of sampling the IMF. 
Thus, we conclude that the clusters associated with clump 4 and 12 are young, likely 4-6~Myr and with masses of $\approx$ 1000-3000~M$_\odot$.
{ Due to the size of the source associated with cloud 4 it is very possible it is a star forming complex and not just a single cluster. This could make    single age  assumed in the SB99 calculations a very strong assumption. }

 \subsection{Star formation across the ALMA field}
The number of star clusters associated with ALMA detected molecular clouds in this study is limited, with only $\sim$ 10\% showing statistically significant presence of a cluster. 
This is despite the observations being sensitive to both young embedded clusters and a more evolved star cluster with the presence of giants, brighter than the sensitivity limit of our survey of Ks$\sim22.5$. 

With the exception of cloud 4, the clumps with NIR counterparts are similar to the general detected CO clumps. Clump 4 is by far the most massive (35.000~M$_\odot$) and has the highest measured velocity dispersion of 2.6 $\mathrm{kms^{-1}}$. 
It is thus an expected clump for the more extended star formation observed in the optical and near-infrared. 
The morphology of the H$\alpha$ and near-infrared emission of cloud 4 suggest older star formation to the north-west of the clump, traced mainly by the H$\alpha$ emission and likely more embedded star formation to the south-east, suggesting an age gradient across the region, from the molecular cloud to the exposed HII region. 
Such a scenario is similar to what is observed in some cases in the outer Milky Way. An example is the H\,{\sc{ii}} region associated with the cluster Dolidze~25 at a distance of 4.5~kpc from the Sun, at a Galactocentric distance of $\approx$ 12~kpc\citep{guarcello,andersen}. 
The central cluster with early-type stars is exposed, partly dispersed and the $\approx$ 3~Myr old cluster has been mostly cleared of material out to a radius of 15~pc and within this region there is substantial H$\alpha$ emission \citep{delgado}. 
Outside there is evidence for substantial molecular material seen as e.g. PAH emission with \textit{Spitzer} \citep{puga09} where some regions are substantially affected by extinction \citep{andersen}. 

The cluster-to-cloud mass is very different for clump 4 and 12. For the former it is low, less than 5\% , whereas for cloud 12 the ratio is almost unity, although the masses of the gas and cluster are uncertain. 
A variable efficiency is in line with theoretical work using, e.g. the STARFORGE simulations \citep{grudic}, depending on the amount of feedback included in the numerical simulations. Other recent high-spatial resolution (2~mpc for a 30-pc box) magnetohydrodynamic numerical simulations of a high mass molecular cloud (10$^4$~M$_\odot$) also show a non-constant efficiency that depends on the feedback mechanisms included in the simulations \citep{2024A&A...682A..76S,2025A&A...698A.119S}.
Further, since the clusters are measured to be several Myr old, it is likely that some of the gas associated with the cluster at birth has been dissociated, or dispersed and is too faint to be detected by ALMA. 

On the basis of the photometry depth, we can estimate the limiting mass of clusters that could be detected in the near-infrared photometry. 
Based on the photometric errors for the point sources in the ALMA field and a nearby comparison field, we reach J and Ks band magnitudes of 23.5 with 0.1 mag photometric error, sources which are easily detected. 
Using this as the limiting magnitude in the mass calculation, we find that for the \citet{kroupa2001} IMF sampled to 100 M$_\odot$, this corresponds to clusters of less than 300 M$_\odot$ seen through an extinction of A$_\mathrm{V}=3$ and ages of a few Myr. 
Thus, the near-infrared observations are sufficiently deep to relatively easily detect even intermediate mass clusters as long as they are unresolved (i.e. less than half an arcsecond or 5 pc).  

We caution that the current study covers one ALMA field of view of $\sim$ 0.45 square arcminutes, corresponding to 
an area of $0.8~kpc^2$, with decreasing sensitivity at the edges of the field. 
Thus, the ALMA surveyed region outside the R$_{25}$ radius is on a size of the smallest scale where global star formation relationships are found \citep[][]{onedera2010,schuba2010,kruijssen2014} and the different time scales and differential drifts between the stellar clusters and CO clumps could potentially create a spatial separation for older clusters (\citep{kodatan}. 
The masses of CO clumps are estimated with the CO(3-2) emission, instead of the commonly-used molecular mass tracers CO(1-0) or CO(2-1).

\section{Conclusions} \label{conc}
We have analyzed deep Gemini Flamingos 2 J and Ks band imaging of the ALMA detected $\mathrm{CO(3-2)}$ clumps in the XUV disk of M83. 
We identified four near-infrared sources spatially coincident with CO clumps. 
On statistical arguments based on the field density of objects, two of those may be chance alignments. 
{ 
Together with Subaru HSC observations, we characterise the cluster mass and ages. One cluster is unresolved with an age of $~$ 4.5~Myr and a mass of 2400~M$_\odot$ and affected by an extinction of A$_\mathrm{V}=3.8$, based on the Starburst99  models with a metallicity of Z=0.008 and an upper mass limit for the \citet{kroupa2001} IMF of 100 M$_\odot$. The other appears to be part of a star-forming complex also visible in H$\alpha$ with a part of the complex almost free of extinction, with a mass and age of 6~Myr and 700~M$_\odot$. The derived masses are higher for an upper mass truncated IMF. For an upper mass limit of 20 M$_\odot$ the cluster masses would be 2.5 times higher and the clusters would be less than $10^7$years. 

 We show that the near-infrared observations are sensitive to clusters with masses of several 100 M$_\odot$ assuming a fully populated IMF, suggesting that massive star formation is relatively rare within the 23 CO clumps identified by \citet{koda22}, and with a total CO cloud mass of 84$\times10^3$ M$_\odot$ the star formation efficiency appears relatively low (less than 10\%\ assuming a fully sampled IMF) in the molecular material in the extended UV disk, but consistent with, e.g. the results from \citet{grudic}}. 

\begin{acknowledgements}
Based on observations obtained at the international Gemini Observatory, a program of NSF NOIRLab, which is managed by the Association of Universities for Research in Astronomy (AURA) under a cooperative agreement with the U.S. National Science Foundation on behalf of the Gemini Observatory partnership: the U.S. National Science Foundation (United States), National Research Council (Canada), Agencia Nacional de Investigaci\'{o}n y Desarrollo (Chile), Ministerio de Ciencia, Tecnolog\'{i}a e Innovaci\'{o}n (Argentina), Minist\'{e}rio da Ci\^{e}ncia, Tecnologia, Inova\c{c}\~{o}es e Comunica\c{c}\~{o}es (Brazil), and Korea Astronomy and Space Science Institute (Republic of Korea). 
J.K. acknowledges support from NSF through grants AST-2006600 and AST-2406608.
This research is based in part on data collected at the Subaru Telescope, which is operated by the National Astronomical Observatory of Japan. We are honored and grateful for the opportunity of observing the Universe from Maunakea, which has the cultural, historical, and natural significance in Hawaii.
\end{acknowledgements}

\bibliographystyle{aa} 
\bibliography{M83} 

\end{document}